\documentclass[lettersize,journal]{IEEEtran} 
\IEEEoverridecommandlockouts

\usepackage{cite}
\usepackage{amsmath,amssymb,amsfonts}
\usepackage{algorithmic}
\usepackage{graphicx}
\usepackage{textcomp}
\usepackage{xcolor}
\def\BibTeX{{\rm B\kern-.05em{\sc i\kern-.025em b}\kern-.08em
    T\kern-.1667em\lower.7ex\hbox{E}\kern-.125emX}}

\usepackage{tcolorbox}
\tcbuselibrary{breakable} 
\newtcolorbox{textbox}[1]{
    sharp corners,
    boxsep=0mm,
    toptitle=2mm,
    lefttitle=0mm,
    colframe=violet!3,
    colback=violet!3,
    title={\rule[-2pt]{4.5pt}{10pt}\hspace*{1.5mm}#1},
    coltitle=violet,
    halign=flush left,
    breakable, 
    before=\par\smallskip\noindent, 
    after=\par\smallskip, 
    parbox=false, 
}

\newtcolorbox{textbox1}[1]{
    sharp corners,
    boxsep=0mm,
    toptitle=2mm,
    lefttitle=0mm,
    colframe=orange!3,
    colback=orange!3,
    title={\rule[-2pt]{4.5pt}{10pt}\hspace*{1.5mm}#1},
    coltitle=orange,
    halign=flush left,
    breakable, 
    before=\par\smallskip\noindent, 
    after=\par\smallskip, 
    parbox=false, 
}

\usepackage{rotating}
\usepackage{booktabs}
\usepackage{multirow}
\usepackage{tabularray}
\usepackage{colortbl}
\usepackage{makecell}
\usepackage{enumitem}
\usepackage{hyperref}
\usepackage{ragged2e}
\usepackage{fontawesome5}
\usepackage{bm}
\usepackage{url}
\usepackage{pdflscape}

\begin{document}

\title{ReqBrain: Task-Specific Instruction Tuning of LLMs for AI-Assisted Requirements Generation}

\author{Mohammad Kasra Habib, Daniel Graziotin,~and~Stefan Wagner
\thanks{M. K. Habib and S. Wagner are with the Chair of Software Engineering, Technical University of Munich, Heilbronn, Germany (e-mail: kasra.habib@tum.de; stefan.wagner@tum.de).\\
ORCID: M. K. Habib \href{https://orcid.org/0000-0002-1272-9873}{0000-0002-1272-9873}, S. Wagner \href{https://orcid.org/0000-0002-5256-8429}{0000-0002-5256-8429}.}
\thanks{D. Graziotin is with the Institute of Information Systems, University of Hohenheim, Stuttgart, Germany (e-mail: graziotin@uni-hohenheim.de).\\
ORCID: \href{https://orcid.org/0000-0002-9107-7681}{0000-0002-9107-7681}.}
}

\maketitle

\begin{abstract}
Requirements elicitation and specification remains a labor-intensive, manual process prone to inconsistencies and gaps, presenting a significant challenge in modern software engineering. 
Emerging studies underscore the potential of employing large language models (LLMs) for automated requirements generation to support requirements elicitation and specification; however, it remains unclear how to implement this effectively. In this work, we introduce ReqBrain, an Al-assisted tool that employs a fine-tuned LLM to generate authentic and adequate software requirements. Software engineers can engage with ReqBrain through chat-based sessions to automatically generate software requirements and categorize them by type. We curated a high-quality dataset of ISO 29148-compliant requirements and fine-tuned five 7B-parameter LLMs to determine the most effective base model for ReqBrain. The top-performing model, Zephyr-7b-beta, achieved 89.30\% Fl using the BERT score and a FRUGAL score of 91.20 in generating authentic and adequate requirements. Human evaluations further confirmed ReqBrain's effectiveness in generating requirements. Our findings suggest that generative Al, when fine-tuned, has the potential to improve requirements elicitation and specification, paving the way for future extensions into areas such as defect identification, test case generation, and agile user story creation.
\end{abstract}


\section{Introduction}\label{sec:intro}
Requirements elicitation and specification is a continuous and fundamental activity in software development \cite{zowghiRequirementsElicitationSurvey2005}. Despite its importance, the process is challenging, manual, and labor-intensive. A significant barrier is tacit knowledge -- information held by a stakeholder but not explicitly shared with the requirements engineer \cite{ferrariAmbiguityResourceDisclose2015, gervasiUnpackingTacitKnowledge2013, distanontEngagementKnowledgeTransfer2012} -- which leads to incomplete requirements, a major pain in requirements engineering \cite{fernandez2017naming}. Clients often struggle to translate objectives into quantifiable requirements, resulting in misunderstandings that may cause deficient or missing critical requirements \cite{ferrariAmbiguityResourceDisclose2015, sutcliffeRequirementsElicitationUnknown2013} and ultimately produce a product lacking essential functionalities \cite{habibDetectingRequirementsSmells2021a}. Natural language requirements lack structured syntax, leading to ambiguity, complexity, and vagueness, which makes understanding difficult for all stakeholders.\cite{mavinEasyApproachRequirements2009, zowghiRequirementsElicitationSurvey2005}.

There is research leveraging AI to explore improving elicitation and specification activities. Earlier approaches, however, were constrained by the limitations of traditional AI techniques: often narrow in focus, lacking cross-domain knowledge, required extensive and carefully structured input, and struggling to capture the complexities of human communication \cite{hovyImportanceModelingSocial2021}. As a result, these tools provided only limited support in real-world scenarios \cite{shakerihosseinabadELICAAutomatedTool2018}. Recent advances in AI enable the understanding of complex context, relationships, and domain knowledge, offering opportunities to proactively support requirements engineers. 

In this study, we focus on an AI-assisted requirements generation approach and concentrate on the early phases of requirements engineering: elicitation and specification, with the potential for future work to expand into a broader range of requirements-related tasks. Whereas \textit{elicitation} implies extracting unexpressed requirements, we define \textit{generation} as the creation of requirements without prior confirmation of their alignment with stakeholder needs.

To support this approach, we propose using large language models (LLMs) to generate software requirements. LLMs trained on large datasets offer a broad cross-domain knowledge base that can support requirements elicitation and specification \cite{devlinBERTPretrainingDeep2019b, radfordImprovingLanguageUnderstanding}. However, general-purpose LLMs might require fine-tuning as they are not specifically designed to generate authentic and adequate requirements, which is essential for overcoming the labor-intensive manual process and ensuring adherence to established requirements engineering standards.

We consider genereated requirements to be \textit{authentic} if they are \textit{indistinguishable} from those written by humans in terms of clarity, coherence, relevance, realism, and implementability. Furthermore, with \textit{adequate}, we refer to four dimensions in AI-generated requirements: \textit{(1) ISO 29148-compliant}~\cite{iso29148}, \textit{(2) consistent} with, \textit{(3) missing} from, and \textit{(4) enhancing the overall completeness} of, a given requirements specification.

With that in mind, we introduce ReqBrain (\textbf{Req}uirements \textbf{Brain}), a fine-tuned LLM and tool to generate authentic and adequate requirements to support the elicitation and specification phases of requirements engineering.

To achieve ReqBrain, we employ task-specific instruction tuning\footnote{\textit{Task-specific instruction tuning} is a supervised fine-tuning method (for more details, see Section \ref{ssec:tsit}). In this paper, we use the term ``\textit{fine-tuning}"\ as a shorter alternative.}. We prefer fine-tuning over prompt engineering due to its ability to improve LLM performance on software engineering tasks and enhance context-specific performance \cite{houLargeLanguageModels2024}. It enables models to internalize task nuances, increasing usability for non-experts \cite{zhangPromptMILBoostingMultiInstance2023} and reducing computational overhead \cite{guPPTPretrainedPrompt2022}. Moreover, it addresses limitations such as prompt length restrictions \cite{guPPTPretrainedPrompt2022}, the risk of knowledge conflict\footnote{Knowledge conflict occurs when the model’s pre-existing training data causes it to interpret a provided instruction or concept differently than intended.} \cite{zhouContextfaithfulPromptingLarge2023}, and reliance on advanced domain expertise \cite{aroraAdvancingRequirementsEngineering2023} to generate requirements.

\textit{\textbf{Our objective is to assess how fine-tuning affects large language models (LLMs) in generating authentic and adequate requirements}}. To achieve our objective for the potential of AI-assisted requirements generation employing ReqBrain, we explore the following research questions:
\vspace{0.1cm}

\noindent \textbf{RQ1: How effectively does a fine-tuned large language model generate authentic requirements?}\\
To address this research question, we split it into the following actionable sub-questions:

\vspace{0.1cm}

\noindent \textit{\textbf{RQ1.1: Which fine-tuned large language model has the highest potential to generate authentic requirements?}}\\
\noindent We benchmark several LLMs, after fine-tuning, using automated NLP metrics. It is crucial to select the model with the highest potential for authentic requirements to reduce the need for exhaustive human evaluation across different models. 

\vspace{0.1cm}

\noindent \textit{\textbf{RQ1.2: Does fine-tuning improve the generation of authentic requirements?}}\\
We aim to understand whether fine-tuning can match or exceed the performance of untuned general-purpose commercial models, in particular ChatGPT-4o, for authentic requirements generation.

\vspace{0.1cm}

\noindent \textit{\textbf{RQ1.3: Are generated requirements perceived as authentic by humans?}}\\
Human evaluators evaluate if ReqBrain generates requirements that are indistinguishable from those authored by humans. This is crucial because achieving human quality standards is fundamental for establishing trustworthiness, user confidence, and integration in development processes.
\vspace{0.1cm}

\noindent \textbf{RQ2: How effectively does a fine-tuned LLM generate adequate requirements?}\\
\noindent Human evaluators assess whether a fine-tuned LLM, \mbox{ReqBrain}, generates adequate requirements. Ensuring that \mbox{ReqBrain} meets the four dimensions of adequate requirements -- ISO 29148-compliant, consistent with, missing from, and enhancing the overall completeness of a given requirements specification -- is critical for generating initial high-quality requirements, saving structuring effort and time to specify requirements unambiguously, preventing costly development issues due to incomplete specifications or identifying potential gaps.

Our work contributes to advancing AI-assisted requirements generation by providing: 
\begin{enumerate}
\renewcommand{\labelenumi}{\arabic{enumi}.}
\item  A novel method and tool for the generation of authentic and adequate requirements.
\item An open \lq instruct\rq\footnote{\lq Instruct\rq\ refers to instructions, with each training instance comprising commands and the expected output, as detailed in Section \ref{sssec:instruct_dataset}.} dataset to support further development and evaluation.
\item Open-source fine-tuned LLMs that enable continual learning and domain adaptation.
\end{enumerate}

\textit{Organization:} The rest of the paper is organized as follows: Section II provides background, Section III related work, Section IV presents requirement generation with ReqBrain, Section V describes the evaluation methodology, Section VI presents results and a discussion, Section VII explores implications, Section VIII discusses threats to validity, and Section IX presents conclusions and future work.

\section{Background}
This section defines key concepts and defines ISO 29148-compliant requirements relevant to our work.

\subsection{Task-Specific Instruction Tuning}\label{ssec:tsit}
Task-specific instruction tuning enhances large language model (LLM) performance on specific tasks by using targeted instructions; in our study, these instructions are about writing requirements. A key benefit of this technique is its ability to reduce data and computational costs while maintaining or improving model effectiveness \cite{chenMaybeOnlyData2023}.

This supervised fine-tuning method represents the instruct dataset as $D = \{(x_1, y_1), (x_2, y_2), \dots, (x_n, y_n)\}$, where each pair $(x_i, y_i)$ consists of an instruction $x_i$ and its corresponding ground truth output $y_i$ (also referred to as \textit{completion}), with $x_i \in X$ and $y_i \in Y$. During the forward pass, the dataset $D$ is input into the language model, producing predicted outputs $\hat{Y} = \{\hat{y}_1, \hat{y}_2, \dots, \hat{y}_n\}$. Then, each pair $(y_i, \hat{y}_i)$ is compared, and gradients are iteratively computed to optimize the model and align its outputs with the ground truth in terms of syntax and semantics. AI-assisted requirements generation can benefit from this fine-tuning method for the automation of various requirements-related tasks.

In our case, $x_i$, $y_i$, and $\hat{y}_i$ are text sequences. For example, $y_i$ is a human-written requirement derived from a real-world project. Based on this requirement,
the corresponding instruction $x_i$ might be: ``Write a functional requirement for a car’s ABS.'' and $\hat{y}_i$ is the model-generated requirement intended to match $y_i$.

\subsection{ISO 29148-Compliant Requirements}\label{ssec:iso_based_requirements}
ISO/IEC/IEEE 29148:2018 \cite{iso29148} provides guidelines for eliciting and specifying high-quality textual requirements in natural language for system and software engineering. We incorporate these guidelines to select high-quality, human-authored requirements to ensure that the training dataset reflects real-world language and the nuanced complexities of industry requirements. 

The subject of ISO-29148-compliant requirements is expansive, and complete coverage of the standard is beyond the scope of this paper. Instead, we limit it to requirements that employ the below-recommended syntaxes and specific signaling keywords, namely, \textit{shall}, \textit{should}, \textit{may}, and \textit{will}.

\vspace{-0.3cm}

{\small
\begin{align}
\textsc{Syntax-1} & : \textit{[Subject]}\textit{[Action]}\textit{[Constraint]} \nonumber \\
\textsc{Syntax-2} & : \textit{[Condition]}\textit{[Subject]}\textit{[Action]}\textit{[Object]}\textit{[Constraint]} \nonumber
\end{align}
}

\section{Related work}

Studies that focus directly on the generation of requirements using LLMs are still limited, leaving a clear research gap: there is a lack of a systematic approach to internalize requirements-engineering-specific knowledge using fine-tuning LLMs to generate authentic and adequate requirements that support simultaneous 
elicitation and specification of user requirements by utilizing broad cross-domain knowledge. Additionally, there is a lack of proper human evaluations to validate the requirements generated by these models.

A related study by Arora, Grundy, and Abdelrazek \cite{aroraAdvancingRequirementsEngineering2023} explores LLM generation across all requirements engineering stages, highlighting use cases in elicitation, specification, and validation, supported by a SWOT analysis and preliminary experiments. They emphasize that prompt design critically impacts output quality in prompt-based LLMs, often leading to inconsistent or overly generic requirements, a limitation experimentally demonstrated in other domains where fine-tuning yields more reliable outputs \cite{tradPromptEngineeringFineTuning2024}.

Similarly, Ronanki, Berger, and Horkoff \cite{ronankiInvestigatingChatGPTsPotential2023} evaluate ChatGPT's potential to generate requirements through controlled experiments. They crafted six elicitation questions and presented them to ChatGPT and human experts. They then compared ChatGPT's outputs with human experts based on abstraction, atomicity, consistency, correctness, unambiguity, understandability, and feasibility. Results show that ChatGPT outperformed human experts in all aspects except unambiguity and feasibility. While LLMs are rich in knowledge, they lack the nuanced, domain-specific understanding needed for authentic and adequate requirement formulation, a limitation that an LLM can learn utilizing fine-tuning \cite{huMitigatingLargeLanguage2024, wangTwostageLLMFinetuning2024}.

While earlier studies have generally explored generating requirements with LLMs, the study by Voria et al. \cite{voriaRECOVERAutomaticRequirements2024a} introduces RECOVER, a pipeline that automatically generates system requirements from stakeholder conversations. The pipeline works by classifying parts of the conversation as requirements segments, cleaning the selected segments, connecting related ideas in conversation, and generating requirements using the LLaMA-2 model. Their results show vulnerability to hallucinations during generation, a known challenge in prompt-driven pipelines, resulting in knowledge conflict or fluent but unfaithful outputs without explicit domain knowledge using fine-tuning \cite{huMitigatingLargeLanguage2024}.

In contrast, AI-assisted requirements generation with \mbox{ReqBrain} addresses these limitations directly. ReqBrain is not tied to a specific technique, an initial set of requirements, stakeholder conversations, interviews, or pre-acquired data. When such data is available, ReqBrain can also be used to extract and generate requirements from it. We fine-tune ReqBrain to generate authentic and adequate requirements using its internal knowledge, and we are the first to investigate the effect of such tuning by employing a systematic approach. LLMs like ReqBrain can encourage dynamic, interactive engagement between requirements engineers and stakeholders, simulating stakeholder perspectives to generate missing requirements and address tacit knowledge gaps while simultaneously specifying the requirements. However, domain experts retain the final decision on accepting, rejecting, or modifying the generated requirements to ensure they align with project-specific ethics, needs, and constraints.

Our contribution directly targets this research gap: the absence of a fine-tuned LLM, a systematic approach for authentic and adequate requirements generation, and the lack of systematic human evaluations to assess such LLMs' output quality.

\section{Requirements Generation with ReqBrain}\label{sec:reqbrain}
ReqBrain is a fine-tuned large language model (LLM) trained to generate authentic and adequate requirements using its internal knowledge to support the AI-assisted requirements generation approach. 

\begin{figure}[h]
\centering
\includegraphics[width=9cm]{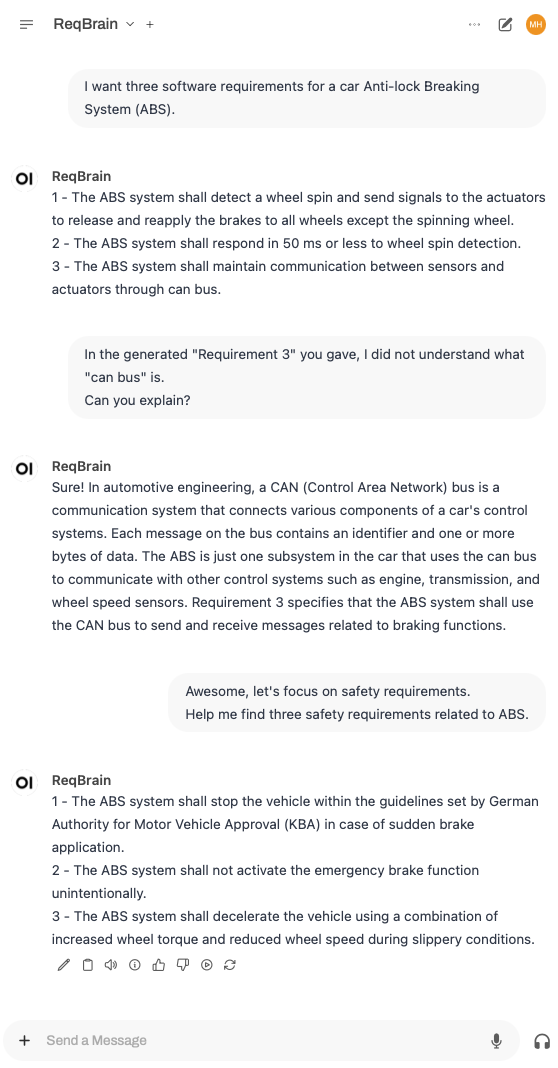}
\caption{AI-assisted generation of software requirements using ReqBrain.}
\label{fig:generation_session}
\end{figure}

Fig.~\ref{fig:generation_session} shows an overview of how a user -- such as requirements engineers, project managers, end users, or customers -- can generate the initial set of requirements through a simple instruction prompt. In this case, a user might start by specifying a target system (e.g., ``Anti-lock brake system'') for which the user wants to generate initial requirements. Then, the user can prompt for further clarification about the generated requirements set, addressing the tacit knowledge by posing additional questions (e.g., ``What is a CAN bus?'') and might continue by concentrating on generating specific requirement types (e.g., ``safety requirements'') to expand the requirements set. This process can continue until the requirements set fully meet the user's objectives.

\begin{figure*}[ht]
\centering
\includegraphics[width=\linewidth]{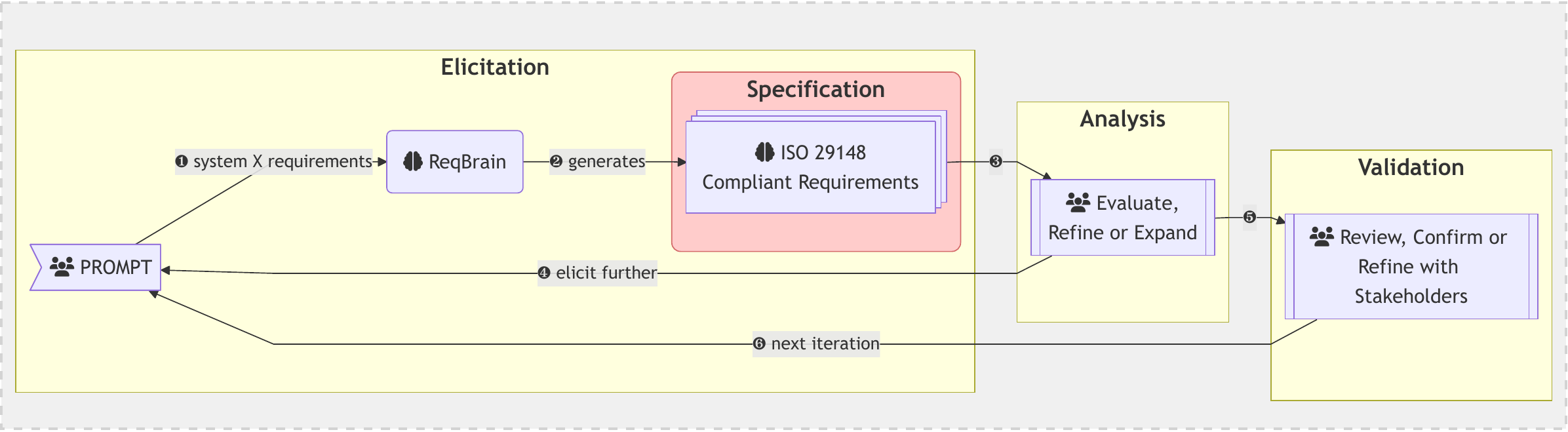}
\caption{AI-assisted requirements generation approach overview, integrating ReqBrain.} 
\label{fig:reqbrain}
\end{figure*}
In addition to that scenario, a user might take a different approach to generate requirements for a target system by inputing a bug or ticket description and prompting ReqBrain to extract core requirements and generate additional ones. Similarly, users can input concrete requirements to ReqBrain and ask for missing ones or enable ReqBrain with retrieval-augmented generation (RAG) to connect it to an internal management system, such as Jira or local git deployments. In Fig. \ref{fig:reqbrain}, we present an overview of ReqBrain's integration into the requirements engineering process based on the software engineering body of knowledge (SWEBOK) activities \cite{SoftwareEngineeringBody}.

To support such capabilities, we incorporate a set of targeted training tasks into the training dataset to improve ReqBrain's performance in generating authentic and adequate requirements:
\begin{itemize}
    \item \textit{How-to? Instructions (INST):} This task is designed to enhance the model's proficiency using ISO 29148-compliance syntax and signaling keywords to generate requirements aligning with industry standards.
    \item \textit{RE-types INST:} This task builds on the previous one to broaden the model's understanding of different requirements classes, such as non-functional security, usability, or functional, to assist the model in accurately differentiating and generating various types of requirements when a user prompts the model.
    \item \textit{Missing INST:} This task simulates scenarios where a set of requirements is incomplete, or a piece of information containing requirements is provided to the model, and the model is tasked to identify missing requirements or extract requirements from the provided information while considering the input information/data.
\end{itemize}

\subsection{Assisting Elicitation and Specification with Generation}
SWEBOK \cite{SoftwareEngineeringBody} defines requirements development in four phases: elicitation, analysis, specification, and validation, which are performed in a sequence iteratively. Our focus is primarily on how generation can assist in the elicitation and specification phases.

As illustrated in Fig. \ref{fig:reqbrain}, generation can be used to simultaneously produce 
requirements compliant to established specification standards, e.g., ISO 29148, during the elicitation phase, effectively merging elicitation with the specification. This reduces manual overhead, shortens iteration cycles, and ensures that early requirements are already in a usable form. Rather than treating elicited data as raw input that needs to be transformed later, generation allows for real-time, semi-automated creation of clear and actionable requirements.

By reducing ambiguity early on, generation shifts the analysis from the second to the third phase, interpreting unclear inputs to make higher-level decisions, such as accepting, rejecting, or modifying generated requirements. 

While other generative approaches are possible, LLMs provide promising support for requirements elicitation and specification. LLMs trained on large datasets offer a broad cross-domain knowledge base that supports requirements elicitation and specification \cite{devlinBERTPretrainingDeep2019b, radfordImprovingLanguageUnderstanding}, where a lack of domain knowledge challenges requirements engineers. Furthermore, LLMs can process large volumes of domain-specific information, such as legacy documentation, Jira, or Git, to generate in-context requirements and save time and effort. 

Additionally, LLMs can encourage requirements engineers and stakeholders to engage in a dynamic, interactive process to shape and refine authentic requirements \cite{aroraAdvancingRequirementsEngineering2023}. They can also simulate stakeholder perspectives \cite{aroraAdvancingRequirementsEngineering2023} to generate missing requirements and address tacit knowledge gaps.

Despite the potential for AI-assisted requirements generation in the elicitation and specification phases, human expert involvement and review remain essential during analysis and validation to ensure alignment with project goals, ethical considerations, emotional intelligence, and contextual understanding.

\subsection{Training Dataset}\label{sssec:instruct_dataset}
To address the absence of a pre-existing dataset for requirements generation, we created an instruct dataset to fill that gap.
As established in Section~\ref{ssec:tsit}, an instruct dataset comprises training records, each represented as a pair $(x_i, y_i)$, where $x_i$ denotes an instruction and $y_i$ its corresponding ground truth output (a human-authored requirement or set of requirements), also referred to as the \textit{completion}. First, we describe the process of collecting and selecting the requirements ($y_i$) and the creation of the instructions ($x_i$) is discussed next.

\subsubsection{Requirements Selection}
We gathered requirements from the Software Requirements Dataset\footnote{The publicly available requirements datasets sourced from within the SwaRD are acknowledged for their contributions: \cite{CoEST, sallam_abualhaija_2021_4471411, ferrariPureDatasetPublic2017, ivanovExtractingSoftwareRequirements2022, knauss_eric_2021_4530183, zain_shaukat_2018_1209601, VHCURES, dalpiazRequirementsClassificationInterpretable2019, hayes_jane_huffman_2018_1223649, oliver_karras_2021_4778899, EPL2}.} (SwaRD), which will soon be released as part of another study. SwaRD consolidates publicly disclosed software requirement artifacts from the internet along with non-disclosed requirements from our industry partners. It includes various types of requirements, such as user stories and acceptance criteria. For this study, we filtered the ISO 29148-compliant requirements from SwaRD.

Although these requirements are labeled as ISO 29148-compliant, we found gaps in their compliance upon closer inspection. The first author manually reviewed and selected the compliant requirements, as outlined in Section \ref{ssec:iso_based_requirements}. While this process is not fully replicable, the selected requirements will be made available within a replication package, allowing interested readers to load the dataset and assess their quality independently.

\subsubsection{Instruction Creation}
To create instructions ($x_i$), we followed established practices and guidelines from the documentation of Hugging Face and OpenAI. For each pair $(x_i, y_i)$, we reviewed the requirement ($y_i$) and created a context that includes supporting information about its intent, class (e.g., functional or non-functional), and ISO 29148-compliance (see Section \ref{ssec:iso_based_requirements}). We then incorporated this context to craft the instruction ($x_i$), paired it with $y_i$, and added the pair to our dataset, $D$. Our templates for writing instructions are given below:

\vspace{-0.3cm}

{\small
\begin{flalign}
& \textit{\textbf{Instruction Template-1}} : [\textit{Context} + \textit{Instruction}] & \nonumber \\
& \textit{\textbf{Instruction Template-2}} : [\textit{Instruction} + \textit{Context}] & \nonumber
\end{flalign}
}

To realize each of the three targeted training tasks, described earlier, we create a corresponding instruction category. First, for instructions in the \textit{How-to? INST} task, we included syntax details such as the correct placement of constraints, conditions, subjects, or signaling keywords, enabling the model to learn the structure of requirements. Next, for instructions in the \textit{RE-types INST} task, we added information about requirement classes using the relabeled PROMISE dataset \cite{dalpiazRequirementsClassificationInterpretable2019}, known for its quality and widespread use in requirement classification studies. Finally, for instructions in the \textit{Missing INST} task, we grouped the selected requirements by their original software projects and split them into two groups: one used in the instruction to simulate an incomplete set of requirements, and the other serving as completion labels.


Table~\ref{tab:inst_comp_trainingset_ex} draws a single instance from each task category from our dataset to illustrate their distinctions and provide an overall understanding.

\begin{table*}
\centering
\caption{Training Pairs from the Instruct Dataset. $^{\ast}$Ellipses are used to condense the text to save space.}
\label{tab:inst_comp_trainingset_ex}
\begin{tblr}{
  width = \linewidth,
  colspec = {Q[58]Q[552]Q[331]},
  row{1} = {c},
  hlines = {0.05em},
  hline{1,5} = {-}{0.08em},
}
\textbf{Task} & \textbf{Instructions} & \textbf{Completion}\\
How to? INST & {Write a requirement in compliance with ISO 29148 recommendations for a well-formed requirement. The desired functionality to be added is to print Construction Junction staff member cards, printing the staff name and unique number in both numeric and barcode formats using the existing company technology (constraints). This functionality must be operational only for registered employees (condition).} & {Once registered online, the system should allow construction junction staff to print out a member card via QuickBooks Point of Sale (POS) containing the member's name and unique member number in both numeric and bar code format.}\\
RE-types INST & {I want my system's customers to have the ability to both purchase and start watching a specific movie within a time frame of fewer than 3 minutes. Can you give me a usability (US) requirement?} & {Customers should be able to purchase and watch a pre-determined movie in under 3 minutes.}\\
Missing INST & {I want to build software for managing project requirements effectively. \\The requirements we have come up with are listed here:\\ - The user is provided the functionality via `` browse"\ mode to manually link low-level requirements to high-level requirements.\\ \textbf{$^{\ast}$\ldots} \\ - The user shall be able to activate the `` save"\ function to save the current data and status of the project.\\ Can you help me complete the missing software requirements?} & {Here are some more software requirements your software might need:\\1 - The `` show"\ function will provide the user with a display of the current file. \\ \textbf{$^{\ast}$\ldots} \\12 - The ``save trace"\ function shall allow the user to save the trace result.}
\end{tblr}
\end{table*}

\subsubsection{Training and Evaluation Sets}\label{ssec:tes}
Our instruction dataset comprises 166 training instances. Together, these instances cover a total of 242 individual requirements. Each instance is a training record that combines an instruction with its corresponding completion. A training record may include multiple requirements, as demonstrated by the \textit{Missing INST} record in Table~\ref{tab:inst_comp_trainingset_ex}.

The dataset is organized into columns containing metadata about the collected requirements, as outlined in Table~\ref{tab:dataset_structure}. Each column serves a specific purpose to ensure that all necessary components are available for efficient model training.

\begin{table}
\centering
\caption{Instruct Dataset Structure, Columns and Description}
\label{tab:dataset_structure}
\resizebox{\linewidth}{!}{%
\begin{tabular}{ll} 
\toprule
\multicolumn{1}{c}{\textbf{Columns}} & \multicolumn{1}{c}{\textbf{Description}} \\ 
\hline
\textbf{REQID\_ex} & An identifier for each requirement \\
\textbf{completion} & Selected requirements from SwaRD ($y_i$).\\
\textbf{instructions} & Created instructions ($x_i$) \\
\textbf{class} & Requirements type information.\\
\textbf{task} & Task: \textit{How-to? INST}; \textit{RE-types INST}; \textit{Missing INST}\\
\textbf{text} & \begin{tabular}[c]{@{}l@{}} Training records: $x_i$ and $y_i$ paired.\end{tabular} \\
\bottomrule
\end{tabular}
}
\end{table}

We applied a stratified split based on targeted task categories, allocating 80\% of the dataset for training and 20\% for evaluation. This method ensures a balanced representation of instruction categories across both sets. 

\subsubsection{Why not a larger training set?}
Creating an extensive training set manually is time-intensive; however, our fine-tuning approach reduces the need for large datasets. For example, Chen et al.~\cite{chenMaybeOnlyData2023} demonstrated that task-specific instruction-tuned models achieve significant performance with only a small fraction of a dataset.

Moreover, fine-tuning an LLM on a large dataset for a specific task can distort its pre-trained weights, leading to catastrophic forgetting and underperformance \cite{kumarFineTuningCanDistort2022, guptaContinualPreTrainingLarge2023}. Several studies, including the influential OpenAI paper on GPT-3 \cite{brownLanguageModelsAre2020} and \cite{tunstallEfficientFewShotLearning2022}, suggest that LLMs require only a few high-quality examples to learn a new task. Therefore, we selected 242 high-quality requirements that best align with our definition of ISO 29148-compliant requirements from, Section \ref{ssec:iso_based_requirements}, instead of creating a large training set.

\subsection{Fine-Tuning}
Training large language models (LLMs) involves two main steps: pre-training and fine-tuning. During pre-training, models are exposed to vast text corpora without task-specific labels or annotations \cite{PreTrainedLanguageModels2023, radfordImprovingLanguageUnderstanding2018, devlinBERTPretrainingDeep2019b}, enabling them to learn general linguistic patterns and structures unsupervised. However, pre-training is costly in terms of computational resources, time, and data requirements, making it less feasible for each new task. Fine-tuning refines a pre-trained model’s representations for a downstream task by updating the pre-trained weights $\Phi_p$ to $\Phi_p + \Delta\Phi_p$, following gradients to maximize the conditional language modeling objective \cite{huLoRALowRankAdaptation2021}.

Fine-tuning is less costly than pre-training because it uses the pre-trained model as a base. However, it has a limitation: learning the complete set of parameters $\Delta\Phi_p$, whose dimension $|\Delta\Phi_p|$ equals $|\Phi_p|$.

Parameter-efficient fine-tuning methods, such as low-rank adaptation (LoRA) \cite{huLoRALowRankAdaptation2021}, address this limitation by reducing the number of parameters updated during training, thereby minimizing the risk of catastrophic forgetting \cite{kirkpatrickOvercomingCatastrophicForgetting2017} -- the loss of pre-trained knowledge during full fine-tuning \cite{chitaleTaskArithmeticLoRA2023, zhaiInvestigatingCatastrophicForgetting2023}. Minimizing catastrophic forgetting is crucial for ReqBrain, as it preserves the base model’s knowledge for use in chat/dialog capabilities to inquire about various aspects of the generated requirements.

LoRA leverages the concept of intrinsic dimension -- the minimum number of dimensions required to represent a matrix’s essential features. In deep learning, training on the intrinsic dimension (i.e., partial training) means updating only a subset, $r$, of $\Phi_p$ for the downstream task \cite{aghajanyanIntrinsicDimensionalityExplains2020}. LoRA achieves this by freezing the pre-trained weights $\Phi_p$, training LoRA weights $\Delta \Phi_l$ for a weight matrix $\Phi_l \in \mathbb{R}^{A \times B}$, and decomposing the weight update matrix into two smaller matrices, as shown in equation 1.

\vspace{-0.3cm}

{\small
\begin{align}
\Delta \Phi_l = \Phi_A \times \Phi_B
\end{align}
}

Where, $\Phi_A \in \mathbb{R}^{A \times r}$ and $\Phi_B \in \mathbb{R}^{r \times B}$, with $r$ representing the intrinsic dimension, a tunable parameter that effectively reduces the number of dimensions. For inference and evaluation, the LoRA weights are added to the original frozen weights $\Phi_p$ at the end of each training round, as shown in equation 2.

\vspace{-0.3cm}

{\small
\begin{align}
h = \Phi_p + \Delta \Phi_l = \Phi_p + AB
\end{align}
}

\subsection{Selecting Models for Fine-Tuning}\label{ssec:model_selection}
For fine-tuning, we focus on open-source models to enable reproducibility and to support organizations in hosting models on their own platforms for privacy. Although tuning commercial models is possible, our key contribution is sharing the dataset and open-source models to enable the models' continual learning, collaboration, and transparency, advancing the AI-assisted requirements generation.

Selecting pre-trained models requires balancing performance and computational resources, which is influenced by model size. Recent studies highlight the effectiveness of 7B LLMs in achieving this balance \cite{creswellSelectionInferenceExploitingLarge2022, laskarBuildingRealWorldMeeting2023}. To establish a baseline and identify the best variant for generating requirements, we initially fine-tuned and compared Falcon-7b-base and the instruct variant, with the instruct variant outperforming the others.

Instruct and chat models interact similarly by providing answers through chat, whereas base models are trained to acquire diverse features, serving as a robust foundation for various tasks. Therefore, we present results for four state-of-the-art open-source instruct or chat models: Llama-2-7b-chat-hf\footnote{As of the training and evaluation period, LLaMA had no instruct-tuned version available.} \cite{LlamaOpenFoundation}, Mistral-7B-Instruct-v0.2 \cite{jiangMistral7B2023}, Zephyr-7b-beta \cite{tunstallZephyrDirectDistillation2023}, Falcon-7b-instruct \cite{falcon40b}, and one base model: Falcon-7b.

\subsection{Fine-Tuning Hyperparameters Selection}
Throughout our experiments, we used the Hugging Face API \cite{HuggingFace} and its models.
For all models, we employed LoRA with $r=64$, as supported by \cite{huLoRALowRankAdaptation2021}, which shows that a low-rank adaptation matrix with $r=64$ effectively captures essential weight update information while ensuring competitive performance and computational efficiency. Based on \cite{hoffmannTrainingComputeOptimalLarge2022}, we opted for a learning rate of $2e-4$ with a cosine scheduler, which balances stability and efficiency by gradually adjusting the learning rate to facilitate stable convergence and mitigate premature stagnation or divergence. For the remaining parameters, we used the original hyperparameters from the base model, as documented in the Hugging Face model's documentation.

\section{Evaluation}
This section outlines the evaluation methodology and study design used to assess ReqBrain’s performance.

\subsection{Study Design}
\textit{\textbf{Our objective is to assess how fine-tuning affects large language models (LLMs) in generating authentic and adequate requirements}}. To achieve our objective, for RQ1.1 and RQ1.2, we used a standard NLP study design. We conducted a between-subjects study design for the remaining research questions to minimize biases such as carryover or learning effects \cite{schuffHowHumanEvaluation2023, greenwaldSubjectsDesignsUse1976}.
By exposing participants to only one condition, the design ensured independent evaluations free from the influence of prior conditions. This independence was crucial for nuanced judgments when comparing ReqBrain-generated requirements against its untuned baseline model and against human-authored requirements or assessing generated requirements for consistency under a single condition \cite{schuffHowHumanEvaluation2023}.

Where applicable, we followed the evaluation guidelines for empirical studies involving LLMs in software engineering by Wagner et al. \cite{wagnerEvaluationGuidelinesEmpirical2025}, reporting model roles, versions, hyperparameters, and hosting details as recommended. Furthermore, all participants provided informed consent prior to their involvement, acknowledging their understanding of the study's purpose, the nature of their participation, and their right to privacy and anonymity.

Participants were asked to bring their laptops to the session, where the study objectives and background information (including knowledge refresher material from ISO 29148) were outlined. They were introduced to three evaluation datasets and their structure. 

At the end of the session, an evaluation package containing essential information and assurances of privacy and anonymity was distributed. Before concluding, participants evaluated a few random requirements from each task to confirm their understanding of the process. Furthermore, all participants provided informed consent prior to their involvement, acknowledging their understanding of the study's purpose, the nature of their participation, and their right to privacy and anonymity.

\subsection{Tasks}
We address our research questions through the following tasks. All tasks, except Task A, are evaluated by human participants.

\subsubsection{Task A}
Within this task, we benchmark the performance of the five fine-tuned LLMs to identify the potential best-performing model in generating authentic requirements for RQ1.1 to reduce exhaustive human evaluation across various models for the subsequent questions. Then, we compare the selected model, ReqBrain, with the untuned ChatGPT-4o-latest\footnote{This is the latest model release of ChatGPT as of 21 December 2024.} to determine whether commercial models designed for general tasks can match or exceed the performance of our fine-tuned model for RQ1.2.

\subsubsection{Task B}
This task compares the requirements generated by ReqBrain with those produced by its untuned baseline model, addressing authenticity in RQ1.3 and the ISO 29148 compliance dimension of adequacy in RQ2. To determine authenticity, participants evaluated how indistinguishable the generated requirements are from those written by humans, focusing on clarity, coherence, relevance, realism, and implementability. For adequacy, we considered the qualities defined for ISO 29148-compliant requirements in Section \ref{ssec:iso_based_requirements}. Human participants evaluated requirements from both models, knowing the set contained a mix of human-authored and AI-generated requirements. We assume a positive fine-tuning effect if participants frequently judge ReqBrain-generated requirements as human-authored.

\subsubsection{Task C}
Building on task B, participants assess authenticity in RQ1.3 and ISO 29148-compliant dimension of adequacy in RQ2 between ReqBrain-generated and human-authored requirements.

\subsubsection{Task D}
We focus on the three remaining dimensions of adequacy in RQ2. We input requirements specifications from real-world projects to ReqBrain and task participants to evaluate whether the generated requirements are consistent with, missing from, and enhance the overall completeness of the given requirements specification.

\subsection{Evaluation Materials}
A comprehensive evaluation set was created for each specific task to assess performance and ensure accurate measurement of outcomes. 

\subsubsection{Benchmark Datasets}\label{sss_em_task_a}
For task A, we generated requirements for each fine-tuned LLM and untuned ChatGPT-4o by inputting the instructions corresponding to the human-authored requirements from our evaluation set detailed in Section \ref{ssec:tes}. Each human-authored requirement is paired with its corresponding LLM-generated requirement for each of the evaluation sets, resulting in a total of six benchmarking datasets.

\subsubsection{ReqBrain vs. Baseline Model Evaluation Dataset}\label{sss_em_task_b}
For task B, we input the instructions from our evaluation set to ReqBrain and its untuned baseline model to generate requirements. To ensure unbiased assessment, the authorship of all generated requirements was anonymized. The requirements were then combined and shuffled before being presented to participants for evaluation.

\subsubsection{ReqBrain vs. Human-Authored Evaluation Dataset}\label{sss_em_task_c}
For Task C, we combined the ReqBrain-generated requirements with their corresponding human-authored counterparts. The requirements were anonymized, shuffled, and presented in a stacked list format for direct comparison.

\subsubsection{ReqBrain Usability Evaluation Dataset}\label{sss_em_task_d}
For Task D, we developed a new evaluation set using requirements from three distinct software projects within KIB\textsuperscript{3} (K\"unstliche Intelligenz in die berufliche Bildung bringen). KIB\textsuperscript{3} is a German AI educational initiative aimed at developing innovative AI tools to support students in their studies. The selected requirements met the following criteria:
\begin{enumerate}
\renewcommand{\labelenumi}{\arabic{enumi}.}
\item Formalized according to ISO 29148 guidelines
\item Elicited from diverse stakeholders, not derived from prior projects or the internet
\item Not published online, reducing the risk of inclusion in any model’s training data
\item Created through a well-documented process
\item Open-source (KIB${^3}$), allowing requirements to be published for transparency and reproducibility
\end{enumerate}

Three software projects were selected: students’ self-evaluation software, adaptation software, and chatbot software. For each project, we created instructions incorporating its requirements and provided them to ReqBrain, which generated additional requirements. The generated requirements were paired with their corresponding instructions and presented to participants for evaluation. In this task, the authorship of the requirements was not concealed, enabling participants to evaluate the generated requirements in full context.

\subsection{Hypotheses, Variables, and Operationalization}
The variables used to measure the authentic and adequate constructs are provided in Table \ref{table:variables}.
To assess how closely AI-generated text aligns with human-authored ground truth in semantics, fluency, coherence, factual accuracy, and originality, we use the established and automated Human Alignment$_{(HA)}$ metrics BERT and FRUGAL. Although the FRUGAL Score is more powerful, BERT is more intuitive; therefore, we computed both. Furthermore, BERT and FRUGAL Scores are learned metrics \cite{zhangBERTScoreEvaluatingText2020, eddineFrugalScoreLearningCheaper2021} that are preferred over traditional metrics such as BLEU, ROUGE, or TER \cite{papineniBleuMethodAutomatic2002, linROUGEPackageAutomatic2004, snoverStudyTranslationEdit2006}, which emphasize surface-form similarity and are often not suitable for computing human alignment \cite{liuExploringCorrelationROUGE2010, elliottComparingAutomaticEvaluation2014, novikovaWhyWeNeed2017}.

\begin{table*}
\centering
\caption{Used Variables for Construct Evaluation.}
\label{table:variables}
\begin{tblr}{
  width = \linewidth,
  colspec = {Q[65]Q[150]Q[537]Q[100]},
  cell{1}{2} = {c},
  cell{1}{3} = {c},
  cell{1}{4} = {c},
  cell{2}{1} = {r=3}{},
  cell{5}{1} = {r=5}{},
  cell{5}{4} = {r=5}{},
  hline{1,10} = {-}{},
  hline{2,5} = {-}{},
}
\textbf{Constructs} & \textbf{Variable} & \textbf{Variable Description} & \textbf{Scale}\\
Authentic & Human Alignment$_{(HA)}$ & BERT Score: measures semantics, fluency, coherence, factual accuracy, and originality with human-authored requirements & Continuous  (0–1)\\
 & Human Alignment$_{(HA)}$ & FRUGAL Score: measures semantics, fluency, coherence, factual accuracy, and originality with human-authored requirements & {Continuous\\(0-100)}\\
 & Perceived Authorship$_{(PA)}$ & Based on the style and content of the requirement, do you believe it was written by a human or generated by AI? & Dichotomous (Human/AI)\\
Adequate & {Written Syntax\\Compliance$_{(WSC)}$} & This requirement is well-structured according to the ISO 29148 recommended syntax. & {Likert\\(1:~Strongly\\Disagree,\\ 5: Strongly\\Agree)}\\
 & {Signaling Keywords\\Compliance$_{(SKC)}$} & This use of signaling keywords to indicate the presence of a requirement is appropriate based on ISO 29148. & \\
 & {Consistent with\\Requirements Set$_{(CRS)}$} & This requirement is consistent with other project requirements. & \\
 & {Identify Missing\\Requirements$_{(IMR)}$} & This requirement accurately reflects the needs that were previously unstated, missing, or overlooked. & \\
 & {Enhancing the \\Overall Completeness$_{(EOC)}$} & Including this requirement would lead to a more complete set of project specifications. & 
\end{tblr}
\end{table*}

We distinguish comparisons (1) between requirements generated by ReqBrain and those produced by its untuned baseline model and (2) between human-authored requirements and those generated by ReqBrain. In the first comparison, failing to reject the null hypothesis indicates no fine-tuning effect. In the second comparison, it suggests a positive effect, as our goal is to achieve human-comparable qualities.

For RQ1.1 and RQ1.2, we used the Human Alignment$_{(HA)}$ variable to evaluate the quality of AI-generated requirements relative to their corresponding human-authored counterparts.

For RQ1.3, we first compare participants’ perceptions and success rates in identifying the requirements generated by \mbox{ReqBrain} against its untuned baseline model as human-authored using the Perceived Authorship$_{(PA)}$ variable and formulate the following hypothesis:

\begin{textbox}{\textit{Hypothesis:}} \justifying
\noindent $H_{0,1}$: The proportion of generated requirements identified as human-authored is independent of whether they were generated by ReqBrain (the fine-tuned model) or its untuned baseline model.

\vspace{0.2cm}

\noindent $H_{a,1}$: The proportion of generated requirements identified as human-authored is not independent of whether they were generated by ReqBrain (the fine-tuned model) or its untuned baseline, with ReqBrain producing a greater proportion.
\end{textbox}

Second, we compare human-authored and ReqBrain-generated requirements using the following hypothesis:

\begin{textbox}{\textit{Hypothesis:}} \justifying
\noindent $H_{0,2}$: Humans do not reliably distinguish between human-authored and ReqBrain-generated requirements in terms of accuracy.

\vspace{0.2cm}

\noindent $H_{a,2}$: Humans reliably distinguish between human-authored and ReqBrain-generated requirements.
\end{textbox}

For the ISO 29148-compliant dimension of adequacy in RQ2, we used the variables Written Syntax Compliance$_{(WSC)}$ and Signaling Keywords Compliance$_{(SKC)}$ to collect participants' responses and formulated the following hypotheses between ReqBrain and its untuned baseline model:

\begin{textbox}{\textit{Hypotheses}} \justifying
\noindent $H_{0,3}$: Requirements generated by ReqBrain do not show greater adherence to ISO 29148 syntax compared to those from its untuned baseline model.

\vspace{0.2cm}

\noindent $H_{a,3}$: Requirements generated by ReqBrain show greater adherence to ISO 29148  syntax compared to those from its untuned baseline model.

\vspace{0.2cm}

\noindent $H_{0,4}$: Requirements generated by ReqBrain do not show greater adherence to ISO 29148 signaling keywords compared to those from its untuned baseline model.

\vspace{0.2cm}

\noindent $H_{a,4}$: Requirements generated by ReqBrain show greater adherence to ISO 29148 signaling keywords compared to those from its untuned baseline model.
\end{textbox}

Next, we compare ReqBrain-generated with human-authored requirements using the following hypotheses:

\begin{textbox}{\textit{Hypotheses:}} \justifying
\noindent $H_{0,5}$: Human-authored and ReqBrain-generated requirements do not differ in their adherence to ISO 29148 written syntax.

\vspace{0.2cm}
 
\noindent $H_{a,5}$: Human-authored and ReqBrain-generated requirements differ in their adherence to ISO 29148 written syntax.

\vspace{0.2cm}

\noindent $H_{0,6}$: Human-authored and ReqBrain-generated requirements do not differ in their adherence to ISO 29148 signaling keywords.

\vspace{0.2cm}

\noindent $H_{a,6}$: Human-authored and ReqBrain-generated requirements differ in their adherence to ISO 29148 signaling keywords.
\end{textbox}

For the remaining dimensions of adequacy in RQ2, we used the variables 
Consistent with Requirements Set$_{(CRS)}$, Missing Requirements$_{(IMR)}$, and Enhancing the Overall Completeness$_{(EOC)}$ to collect participants responses. Responses $\leq 3$ indicate a range from neutral to strongly disagree on our selected Likert scale.

\begin{textbox}{\textit{Hypotheses:}} \justifying
\noindent $H_{0,7}$: The median rating ($M$) for the Consistent with Requirements Set$_{(CRS)}$ is  $\leq 3$.

\vspace{0.2cm}

\noindent $H_{a,7}$: The median rating for Consistent with Requirements Set$_{(CRS)}$ is $> 3$.

\vspace{0.2cm}

\noindent $H_{0,8}$: The median rating ($M$) for the Identify Missing Requirements$_{(IMR)}$ is $\leq 3$.

\vspace{0.2cm}

\noindent $H_{a,8}$: The median rating for Identify Missing Requirements$_{(IMR)}$ is $> 3$.

\vspace{0.2cm}

\noindent $H_{0,9}$: The median rating ($M$) for the Enhancing the Overall Completeness$_{(EOC)}$ is $\leq 3$.

\vspace{0.2cm}

\noindent $H_{a,9}$: The median rating for Enhancing the Overall Completeness$_{(EOC)}$ is $> 3$
\end{textbox}

\subsection{Analysis Procedure}
First, we analyze RQ1.1 and RQ1.2 using NLP metrics to measure the similarity between requirements. Then, we outline the evaluation process for the remaining research questions.

\subsubsection{NLP Metrics Analysis Procedure}
We compute the pairwise similarity between the ReqBrain-generated and human-authored ground truth requirements (see Section \ref{sss_em_task_a} for the evaluation set setup) for RQ1.1 and RQ1.2.

\paragraph{The BERT Score}
BERT score is a learned evaluation metric for text generation that utilizes contextualized embeddings from BERT \cite{devlinBERTPretrainingDeep2019b}. It computes the cosine similarity between token embeddings of a human-authored reference $x = {x_1, x_2, \ldots, x_n}$ and an AI-generated equivalent $\hat{x} = {\hat{x}_1, \hat{x}_2, \ldots, \hat{x}_m}$. The precision and recall are computed as:

{\small
\begin{align}
R = & \frac{1}{|x|}{\sum_{x_i \in x} \max_{\hat{x}j \in \hat{x}} x_i^\top \hat{x}j}\ \textit{and}\ P = \frac{1}{|\hat{x}|}{\sum{x_i \in x} \max{\hat{x}_j \in \hat{x}} x_i^\top \hat{x}_j}
\end{align}
}

The F1 score is derived from these values. Unlike traditional $n$-gram metrics, contextualized embeddings capture word meaning, synonyms, context, and grammar \cite{devlinBERTPretrainingDeep2019b}.

\paragraph{The FRUGAL Score}
The FRUGAL score is similar to the BERT score but is faster and lighter, and in some cases, outperforms the BERT score \cite{eddineFrugalScoreLearningCheaper2021}. Its training involves generating a synthetic dataset by pairing sequences annotated with costly metrics aligned with human judgment, followed by pre-training a miniature language model\footnote{Is a downscaled larger model that maintains the original performance or comes close to it \cite{turcWellReadStudentsLearn2019a}.} on this dataset to learn the mapping of costly metrics and similarity functions.

\subsubsection{Human Evaluation Analysis Procedure}
For RQ1.3 and the four dimensions in RQ2, we used descriptive statistics to summarize sample characteristics and inferential statistics to test the hypotheses.

For all samples used in RQ2, we calculated the mean ($\tilde{x}$), standard deviation ($s$), and median ($M$). Although we report $\tilde{x}$ and $s$, hypothesis testing relies on the median ($M$), which is more appropriate for ordinal data \cite{leechCallGreaterUse2002, schuffHowHumanEvaluation2023}.

Due to the nature of the data, non-parametric tests were employed to evaluate all the hypotheses. Non-parametric tests are robust for ordinal data and do not assume normality or equal intervals \cite{schuffHowHumanEvaluation2023, arcuriPracticalGuideUsing2011, varghaCritiqueImprovementCL2000}. For all tests, a significance level of $\alpha = .05$ was used to determine statistical significance, and 95\% confidence intervals were reported for effect size estimates.

For RQ1.3, descriptive statistics include success and failure counts and success proportions. A right-tailed Fisher’s Exact test was used to test $H_{a,1}$, comparing the proportions of requirements identified as human-authored between ReqBrain and its untuned baseline model. An odds ratio was calculated as the effect size.

For identifying authorship between ReqBrain-generated and human-authored requirements, a contingency table and expected frequencies were computed for both samples employing the Chi-square test ($\chi^2$) to evaluate $H_{a,2}$. Overall human precision in identifying authorship between ReqBrain-generated and human-authored requirements was also calculated with confidence intervals.

For the ISO 29148-compliant dimension of adequacy in RQ2, four hypotheses are tested. Two hypotheses ($H_{a,3}$ and $H_{a,4}$) compare ISO 29148-compliance between ReqBrain-generated and its untuned baseline model using right-tailed Mann-Whitney U tests. The remaining two ($H_{a,5}$ and $H_{a,6}$) compare ReqBrain-generated requirements with human-authored ones using two-tailed Mann-Whitney U tests to evaluate equivalence. The effect size for all Mann-Whitney U tests is quantified using Vargha and Delaney’s A-statistic \cite{varghaCritiqueImprovementCL2000}.

For the remaining dimensions of adequacy in RQ2, three one-sample Wilcoxon signed-rank tests were conducted, one for each hypothesis ($H_{a,7}$, $H_{a,8}$, and $H_{a,9}$), with the rank biserial ($r$) effect size. Additionally, to account for multiple tests for RQ1.3 and each dimension in RQ2, we calculate and report adjusted p-values using the Holm-Bonferroni method \cite{holmSimpleSequentiallyRejective1979}.

\subsection{Sample Size and Participants}
We performed an a-priori power analysis to determine the sample size for different evaluations. Following Dyb\aa\ et al.~\cite{dybaSystematicReviewStatistical2006}, we conducted power analysis for the non-parametric tests using their analogous parametric tests. We used the conventional $\alpha = 0.05$, power $= 0.8$, and a recommended effect size $\beta = 0.5$ for software engineering studies \cite{dybaSystematicReviewStatistical2006}. An optimal sample size of 64 requirements was calculated for two-tailed Mann-Whitney U tests and 51 for its one-tailed tests using a two-tailed t-test and one-tailed t-test, respectively. For one-sample, one-tailed Wilcoxon signed-rank tests, an optimal sample size of 26 was determined using a one-sample, one-tailed t-test, and for Fisher's exact, we used Chi-square to calculate an optimal sample size of 32. The first part of our study design (see Section~\ref{ssec:tes}) resulted in sample sizes two to three times larger.

Four experienced participants evaluated a total of 672 distinct requirements for different tasks. Two had 1--3 years of work experience, and the other had 4--6 years in software and requirements engineering. Each participant also held at least a bachelor's degree in software engineering. In terms of familiarity with AI content, two were “Very familiar,” one was “Somewhat familiar,” and one was “Moderately familiar.” Regarding the use of generative AI tools like ChatGPT, two used them “Sometimes,” one answered “Yes,” and one responded “No.”

Table~\ref{table:summary_ev} presents a comprehensive overview of the main points in this section.
\begin{table*}[h]\huge
\centering
\caption{RQ Mapping to Hypotheses, Evaluation Materials, Variables, Statistical Tests, Directionality, and Compared Samples. Abbreviation: $M_h$, Hypothesized Median}
\label{table:summary_ev}
\resizebox{\linewidth}{!}{%
\begin{tblr}{
  row{1} = {c},
  cell{2}{1} = {c},
  cell{2}{2} = {r=2}{},
  cell{2}{3} = {r=2}{},
  cell{2}{4} = {r=2}{},
  cell{2}{5} = {r=2}{},
  cell{2}{6} = {r=2}{c},
  cell{3}{1} = {c},
  cell{4}{1} = {c},
  cell{4}{4} = {r=2}{},
  cell{4}{6} = {c},
  cell{5}{1} = {c},
  cell{5}{6} = {c},
  cell{6}{1} = {r=3}{c},
  cell{6}{4} = {r=2}{},
  cell{6}{6} = {c},
  cell{7}{6} = {c},
  cell{8}{6} = {c},
  hline{1-2,9} = {-}{},
}
\textbf{RQ} & \textbf{Hypotheses} & \textbf{Evaluation Materials} & \textbf{Variables} & \textbf{Statistical Test} & \textbf{Directionality} & \textbf{Compared Samples}\\
\textbf{RQ1.1} & n/a & Benchmark Datasets & Human Alignment$_{(HA)}$ & {BERT Score;\\FRUGAL Scores} & n/a & Human vs. ReqBrain\\
\textbf{RQ1.2} &  &  &  &  &  & {Human vs. ReqBrain\\Human vs. ChatGPT-4o}\\
\textbf{RQ1.3} & {$H_0,1$,\\$ H_a,1$} & {ReqBrain vs. its Baseline Model\\~Evaluation Dataset} & Perceived Authorship$_{(PA)}$ & Fisher's Exact & Right-tailed & {ReqBrain vs. \\its Baseline Model}\\
\textbf{RQ1.3} & {$H_0,2$,\\$H_a,2$} & {ReqBrain vs. Human-Authored \\Evaluation Dataset} &  & Chi-square & n/a & Human vs. ReqBrain\\
\textbf{RQ2} & {$H_0,3-4$,\\$H_a,3-4$} & {ReqBrain vs. Baseline Model \\Evaluation Dataset} & {Written Syntax Compliance$_{(WSC)}$,\\Signaling Keywords Compliance$_{(SKC)}$} & Mann-Whitney U & Right-tailed & {ReqBrain vs. \\its Baseline Model}\\
 & {$H_0,5-6$,\\$H_a,5-6$} & {ReqBrain vs. Human-Authored\\~Evaluation Dataset} &  & Mann-Whitney U & Two-tailed & Human vs. ReqBrain\\
 & {$H_0,7-9$,\\$H_a,7-9$} & {ReqBrain Usability\\~Evaluation Dataset} & {Consistent with Requirements Set$_{(CRS)}$,\\Identify Missing Requirements$_{(IMR)}$,\\Enhancing the Overall Completeness$_{(EOC)}$} & {Wilcoxon \\Signed-rank} & Right-tailed & {ReqBrain vs.~\\$M_h > 3$}
\end{tblr}
}
\end{table*}

\section{Results and Discussion}\label{sec:results}
We first present the results corresponding to each research question, followed by a brief discussion and then a summary of findings.

\subsection{RQ1 Generating Authentic Requirements}
\subsubsection{RQ1.1 Benchmarking the Fine-tuned Models} Table \ref{table:rq1_1} presents the performance metrics of five fine-tuned large language models. Zephyr-7b-beta outperforms all other models on both metrics.

\begin{table}[h ]
\centering
\caption{\textnormal{Human Alignment$_{(HA)}$} Results: Performance Metrics for Five Fine-Tuned LLMs. Abbreviations: P, Precision; R, Recall.}
\label{table:rq1_1}
\resizebox{\linewidth}{!}{%
\begin{tabular}{lcccc} 
\hline
\multicolumn{1}{c}{\multirow{2}{*}{\textbf{Models}}} & \multicolumn{3}{c}{\textbf{BERT Score}} & \multirow{2}{*}{\textbf{FRUGAL Score}} \\ 
\cline{2-4}
\multicolumn{1}{c}{} & \textbf{P} & \textbf{R} & \textbf{F1} &  \\ 
\hline
\textbf{Zephyr-7b-beta} & \textbf{0.89} & \textbf{0.89} & \textbf{0.89} & \textbf{0.91} \\
\textbf{Mistral-7B-Instruct-v0.2} & 0.84 & \textbf{0.89} & 0.86 & 0.88 \\
\textbf{Falcon-7b} & 0.80 & 0.82 & 0.81 & 0.88 \\
\textbf{Falcon-7b-instruct} & 0.85 & 0.88 & 0.86 & 0.88 \\
\textbf{\textbf{Llama-2-7b-chat-hf}} & \multicolumn{1}{l}{0.81} & \multicolumn{1}{l}{0.85} & \multicolumn{1}{l}{0.83} & 0.88 \\
\hline
\end{tabular}
}
\end{table}

\begin{figure}[h]
\centering
\includegraphics[width=1.05\columnwidth]{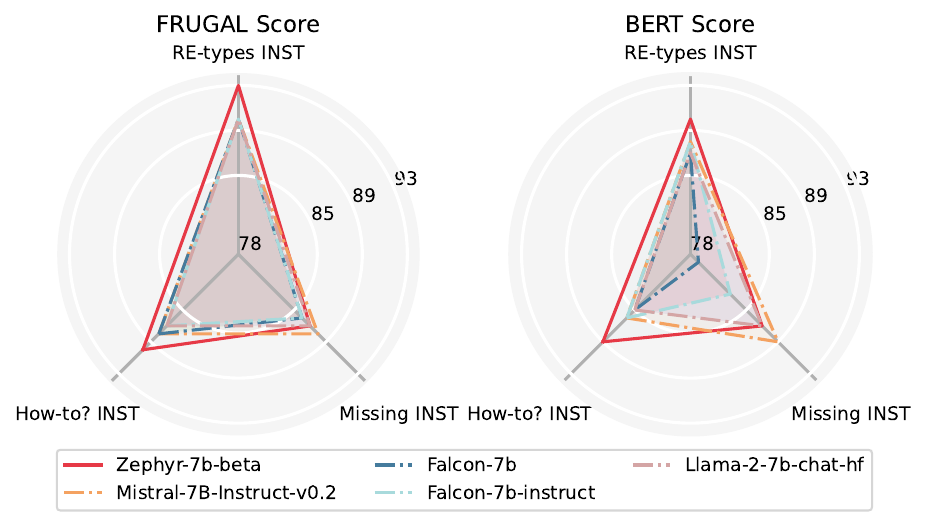}
\caption{Performance metrics across three task categories (see Section \ref{sssec:instruct_dataset}).}
\label{fig:spider_chart}
\end{figure}

Figure \ref{fig:spider_chart} illustrates the performance of the models across the three instruction categories described in Section \ref{sssec:instruct_dataset} using FRUGAL and BERT scores. Both scores show that Mistral-7B-Instruct-v0.2 performs slightly better than Zephyr-7b-beta in the \textit{Missing INST} task. This may stem from its architectures and training data, which likely include a broader range of similar tasks.

Nevertheless, Zephyr-7b-beta records higher aggregate scores across all three task categories. Hence, we identified Zephyr-7b-beta as the most effective model for generating authentic requirements.

\subsubsection{RQ1.2 Benchmarking ReqBrain against ChatGPT-4o} Table \ref{table:rq1_2} summarizes performance metrics for ReqBrain, our best-performing fine-tuned model, and untuned ChatGPT-4o. The metrics show that ReqBrain outperforms the untuned ChatGPT-4o in generating authentic requirements.

\begin{table}[h]
\centering
\caption{\textnormal{Human Alignment$_{(HA)}$} Results: Performance Metrics for ReqBrain vs. ChatGPT-4o. Abbreviations: P, Precision; R, Recall.}
\label{table:rq1_2}
\resizebox{0.9\columnwidth}{!}{
\begin{tabular}{ccccc} 
\hline
\multirow{2}{*}{\textbf{Models}} & \multicolumn{3}{c}{\textbf{BERT Score}} & \multirow{2}{*}{\textbf{FRUGAL Score}} \\ 
\cline{2-4}
 & \textbf{P} & \textbf{R} & \textbf{F1} &  \\ 
\hline
\multicolumn{1}{l}{\textbf{ReqBrain}} & \textbf{0.89} & \textbf{0.89} & \textbf{0.89} & \textbf{0.91} \\
\multicolumn{1}{l}{\textbf{ChatGPT-4o}} & 0.81 & 0.88 & 0.84 & 0.86 \\
\hline
\end{tabular}
}
\end{table}
Although the comparison with untuned ChatGPT-4o might seem unfair, it underscores the importance of fine-tuning LLMs for requirements elicitation tasks. ChatGPT-4o, with its larger parameter count, might surpass ReqBrain (which has 7 billion parameters) in performance if fine-tuned. 

\subsubsection{RQ1.3 ReqBrain vs. its Untuned Baseline Model} Table \ref{tab:test_results_rq2} provides results to assess the perceived human authorship of requirements generated by ReqBrain and its untuned baseline model.

\begin{table}[h]
\centering
\caption{\textnormal{Perceived Authorship$_{(PA)}$} Results: Perceived Human-Likeness Between ReqBrain and its Untuned Baseline Model.}
\label{tab:test_results_rq2}
\resizebox{\linewidth}{!}{%
\begin{tblr}{
  column{2} = {c},
  column{3} = {c},
  cell{6}{2} = {c=2}{},
  cell{7}{2} = {c=2}{},
  cell{8}{2} = {c=2}{},
  hline{1-2,6,9} = {-}{},
}
\textbf{Attribute} & \textbf{ReqBrain} & \textbf{Untuned Baseline}\\
Sample Size & $n_1$: 136 & $n_2$: 136\\
Human-Classified Count & 65 & 12\\
AI-Classified Count & 71 & 124\\
Human-identified Proportions & 0.47 & 0.08\\
Odds Ratio & 9.46 (95\% $CI = [4.79, 18.70]$) & \\
$p$-Value &  $<.001$ & \\
$Adj.\ p$ &  $<.001$ & 
\end{tblr}
}
\end{table}

For ReqBrain, 47.8\% of the generated requirements are identified as human-authored, compared to only 8.8\% for the untuned baseline model. The right-tailed Fisher’s Exact test produced a p-value $< 0.001$, providing strong evidence in favor of the alternative hypothesis ($H_{a,1}$). The odds ratio of 9.46 indicates that the odds of fine-tuned model outputs being perceived as authentic are approximately 9.5 times higher than those of the baseline model.

\subsubsection{RQ1.3 ReqBrain vs. Human Authors} In Table \ref{tab:test_results_rq3}, we summarize the results for this comparison. The Chi-square test yielded $\chi^2(1; N = 272) = 0.01475694$ and $p = 0.90331$, with an odds ratio of 1.06, providing no evidence to support the alternative hypothesis ($H_{a,2})$. Furthermore, the classification precision is 50.7\%. 
The results suggest that ReqBrain-generated requirements are perceived as authentic by humans, as evaluators could not reliably distinguish between them and those authored by humans.

\begin{table}[h]
\centering
\caption{\textnormal{Perceived Authorship$_{(PA)}$} Results: Human Ability to Distinguish ReqBrain-Generated Requirements from Human-Authored Requirements}
\label{tab:test_results_rq3}
\resizebox{\linewidth}{!}{%
\begin{tblr}{
  column{3} = {c},
  cell{1}{2} = {c},
  cell{2}{2} = {c},
  cell{3}{2} = {c=2}{c},
  cell{4}{2} = {c=2}{c},
  cell{5}{2} = {c=2}{c},
  cell{6}{2} = {c=2}{c},
  cell{7}{2} = {c=2}{c},
  cell{8}{2} = {c=2}{},
  cell{9}{2} = {c=2}{c},
  hline{1-2,5,10} = {-}{},
}
\textbf{Attribute} & \textbf{Human} & \textbf{ReqBrain}\\
Sample Size & $n_1$: 136 & $n_2$: 136\\
Contingency Frequencies & $[[65, 71] [63, 73]]$ & \\
Expected Frequencies & $[[64, 72] [64, 72]]$ & \\
$\chi^2$ Statistics & 0.01475694 & \\
$p$-Value & $0.90331$ & \\
$Adj. p$ & $0.90331$ & \\
Odds Ratio & 1.06 (95\%~$CI = [0.64, 1.75]$) & \\
Human Identification Precision & 50.7\% (95\% $CI = [44\%, 56\%]$) & 
\end{tblr}
}
\end{table}

\begin{textbox1}{\textit{Findings RQ1:}} \justifying
Our study shows that fine-tuning LLMs effectively generates authentic requirements indistinguishable from those authored by humans. Human evaluators could not reliably differentiate between human-authored and fine-tuned LLM-generated requirements (ReqBrain), indicating that its outputs meet human quality standards.
\end{textbox1}

\subsection{RQ2 Generating Adequate Requirements}
\subsubsection{Comparing ReqBrain and its Untuned Baseline Model on ISO 29148} Table \ref{tab:test_results_rq4_syntax_sig} provides a summary of the results. For written syntax, ReqBrain achieved a median rating of 4 compared to 2 for the untuned baseline model. The right-tailed Mann-Whitney U test yielded $U = 14203.5$ and $p < .001$, with large effect size $A_{12} = 0.76$, indicating strong evidence in favor of the alternative hypothesis ($H_{a,3}$). For signaling keywords, ReqBrain also achieved a median rating of 4 compared to 2 for the untuned baseline model, with $U = 13766.0$ and $p < .001$, and large effect size $A_{12} = 0.74$, supporting the alternative hypothesis ($H_{a,4}$).

The results suggest that a fine-tuned LLM significantly outperforms its untuned baseline model in generating ISO 29148-compliant requirements.

\begin{table*}[ht]
\centering
\caption{ISO 29148-Compliant Dimensions of Adequate: ReqBrain vs. its Untuned Baseline Model Abbreviations: $N$, sample size; $M$, median; $\tilde{x}$, mean; $p$, p-value; $Adj.\ p$, Holm-Bonferroni adjusted p-value; $U$, Mann-Whitney U; $A_{12}$, Vargha and Delaney’s effect size.}
\label{tab:test_results_rq4_syntax_sig}
\resizebox{\linewidth}{!}{%
\begin{tblr}{
  column{1} = {c},
  column{3} = {c},
  column{4} = {c},
  column{5} = {c},
  column{6} = {c},
  column{7} = {c},
  column{8} = {c},
  column{10} = {c},
  column{11} = {c},
  column{12} = {c},
  cell{1}{2} = {c},
  cell{2}{1} = {r=2}{l},
  cell{2}{3} = {r=2}{},
  cell{2}{8} = {r=2}{},
  cell{2}{9} = {r=2}{},
  cell{2}{10} = {r=2}{},
  cell{2}{11} = {r=2}{},
  cell{2}{12} = {r=2}{},
  cell{4}{1} = {r=2}{l},
  cell{4}{3} = {r=2}{},
  cell{4}{8} = {r=2}{},
  cell{4}{9} = {r=2}{},
  cell{4}{10} = {r=2}{},
  cell{4}{11} = {r=2}{},
  cell{4}{12} = {r=2}{},
  hline{1,6} = {-}{0.08em},
  hline{2} = {-}{},
}
\textbf{Variables} & \textbf{Author} & \textbf{Hypotheses} & $N$ & $M$ & $\tilde{x}$ & $s$ & \textbf{$p$} & \textbf{$Adj.\ p$} & \textbf{$U$} & $A_{12}$ & \textbf{95\%$ A_{12}$ CI}\\
{Written Syntax\\Compliance$_{(WSC)}$} & ReqBrain & {$H_{0,3}$, $H_{a,3}$} & $n_1$: 136 & 4 & 3.78 & 1.26 & $< .001$ & $< .001$ & 14203.5 & 0.76 & $[0.71, 0.82]$\\
 & Untuned Baseline &  & $n_2$: 136 & 2 & 2.316 & 1.45 &  &  &  &  & \\
{Signaling Keywords\\Compliance$_{(SKC)}$} & ReqBrain & {$H_{0,4}$, $H_{a,4}$} & $n_1$: 136 & 4 & 3.90 & 1.27 & $< .001$ & $< .001$ & 13766.0 & 0.74 & $[0.68, 0.80]$\\
 & Untuned Baseline &  & $n_2$: 136 & 2 & 2.64 & 1.38 &  &  &  &  & 
\end{tblr}
}
\end{table*}

\vspace{0.1cm}

\subsubsection{Comparing ReqBrain and Humans on ISO 29148} Table \ref{tab:test_results_rq5_syntax_sig} provides a summary of the results. For written syntax, both groups achieved a median rating of 4. The Mann-Whitney U test revealed no significant difference with $U = 10118.5$ and $p = 0.15521$, with an effect size of $A_{12} = 0.54$, thereby providing no support for the alternative hypothesis ($H_{a,5}$). For signaling keywords, both groups recorded a median rating of 4. However, the Mann-Whitney U test revealed a statistically significant difference with $U = 10482.0$ and $p = 0.04068$, supporting the alternative hypothesis ($H_{a,6}$) and a small effect size $A_{12} = 0.56$. Nevertheless, observing its $Adj.\ p = 0.12204$ suggests rejecting the alternative hypothesis; note that for all other hypotheses, the adjusted $p$-values confirm the unadjusted $p$-values.

The findings suggest that fine-tuned LLM produces requirements that are comparable to those authored by humans in terms of appropriate use of syntax and signaling keyword usage.

\begin{table*}
\centering
\caption{ISO 29148-Compliant Dimensions of Adequate: ReqBrain vs. Human Authors. Abbreviations: $N$, sample size; $M$, median; $\tilde{x}$, mean; $p$, p-value; $Adj.\ p$, Holm-Bonferroni adjusted p-value; $U$, Mann-Whitney U; $A_{12}$, Vargha and Delaney’s effect size.}
\label{tab:test_results_rq5_syntax_sig}
\resizebox{\linewidth}{!}{%
\begin{tblr}{
  column{1} = {c},
  column{3} = {c},
  column{4} = {c},
  column{5} = {c},
  column{6} = {c},
  column{7} = {c},
  column{8} = {c},
  column{10} = {c},
  column{11} = {c},
  column{12} = {c},
  cell{1}{2} = {c},
  cell{2}{1} = {r=2}{l},
  cell{2}{3} = {r=2}{},
  cell{2}{8} = {r=2}{},
  cell{2}{9} = {r=2}{},
  cell{2}{10} = {r=2}{},
  cell{2}{11} = {r=2}{},
  cell{2}{12} = {r=2}{},
  cell{4}{1} = {r=2}{l},
  cell{4}{3} = {r=2}{},
  cell{4}{8} = {r=2}{},
  cell{4}{9} = {r=2}{},
  cell{4}{10} = {r=2}{},
  cell{4}{11} = {r=2}{},
  cell{4}{12} = {r=2}{},
  hline{1,6} = {-}{0.08em},
  hline{2} = {-}{},
}
\textbf{Variables} & \textbf{Author} & \textbf{Hypotheses} & $N$ & $M$ & $\tilde{x}$ & $s$ & \textbf{$p$} & \textbf{$Adj.\ p$} & \textbf{$U$} & $A_{12}$ & \textbf{95\%$ A_{12}$ CI}\\
{Written Syntax\\Compliance$_{(WSC)}$} & Human & {$H_{0,5}$, $H_{a,5}$} & $n_1$: 136 & 4 & 3.80 & 1.20 & $0.15521$ & $0.31043$ & 10118.5 & 0.54 & $[0.48, 0.61]$\\
 & ReqBrain &  & $n_2$: 136 & 4 & 3.58 & 1.30 &  &  &  &  & \\
{Signaling Keywords\\Compliance$_{(SKC)}$} & Human & {$H_{0,6}$, $H_{a,6}$} & $n_1$: 136 & 4 & 4.11 & 0.96 & $0.04068$ & $0.12204$ & 10482.0 & 0.56 & $[0.50. 0.63]$\\
 & ReqBrain &  & $n_2$: 136 & 4 & 3.83 & 1.14 &  &  &  &  & 
\end{tblr}
}
\end{table*}

\vspace{0.1cm}
\subsubsection{Evaluating ReqBrain on the Remaining Dimensions} Table \ref{tab:wilcoxon_results_transposed} summarizes the results for these three dimensions. For all three dimensions, the median rating was $M = 4$. 

For \textit{consistent with} dimension, the right-tailed Wilcoxon Signed Rank test yielded a statistically significant median rating greater than 3 with $W = 7470.5$ and $p < .001$ with a large effect size $r = 0.82$ supporting the alternative hypothesis ($H_{a,7}$). 

For \textit{missing from} dimension, the right-tailed Wilcoxon Signed Rank test showed a statistically significant median rating greater than 3 with $W = 7035.0$ and $p < .001$, and a large effect size $r = 0.72$. These results support the alternative hypothesis ($H_{a,8}$). 

For \textit{enhancing the overall completeness} dimension, the right-tailed Wilcoxon Signed Rank test with $W = 6378.0$ and $p < .001$ confirms the rejection of the null hypothesis in favor of the alternative ($H_{a,9}$), with a large effect size $r = 0.58$. 

In summary, all three dimensions confirm that ReqBrain is effective in generating requirements that are consistent with, missing from, and enhancing the overall completeness of a given specification.

\begin{table*}\small
\centering
\caption{ReqBrain Effectiveness on Remaining Dimensions of Adequate. Abbreviations: $n$, sample size; $M$, median; $\tilde{x}$, mean; $s$, standard deviation; $p$, p-value; $Adj.\ p$, Holm-Bonferroni adjusted p-value; $r$, Rank Biserial effect size.}
\label{tab:wilcoxon_results_transposed}
\resizebox{\linewidth}{!}{%
\begin{tblr}{
  column{even} = {c},
  column{3} = {c},
  column{5} = {c},
  column{7} = {c},
  column{11} = {c},
  cell{1}{1} = {c},
  hline{1,5} = {-}{0.08em},
  hline{2} = {-}{},
}
\textbf{Variables} & \textbf{Hypotheses} & \textbf{$n$} & \textbf{$M$} & \textbf{$\tilde{x}$} & $s$ & \textbf{Wilcoxon Signed-Rank} & \textbf{$p$} & \textbf{$Adj.\ p$} & \textbf{$r$} & \textbf{95\% $r$ CI}\\
{Consistent with\\Requirements Set$_{(CRS)}$} & {$H_{0,7}$, $H_{a,7}$} & 128 & 4 & 4.0 & 1.02 & 7470.5 & $< .001$ & $< .001$ & 0.82 & $[0.72, 0.94]$\\
{Identify Missing\\Requirements$_{(IMR)}$} & {$H_{0,8}$, $H_{a,8}$} & 128 & 4 & 3.86 & 1.06 & 7035.0 & $< .001$ & $< .001$ & 0.71 & $[0.58, 0.85]$\\
{Enhancing the\\Overall Completeness$_{(EOC)}$} & {$H_{0,9}$, $H_{a,9}$} & 128 & 4 & 3.63 & 1.13 & 6378.0 & $< .001$ & $< .001$ & 0.57 & $[0.44, 0.74]$
\end{tblr}
}
\end{table*}

\begin{textbox1}{\textit{Findings RQ2:}} \justifying
Our results indicate that a fine-tuned LLM generates adequate requirements, thereby validating its effectiveness in generating high-quality requirements.
\end{textbox1}

\section{Implications for Research and Practice}
Below, we discuss how ReqBrain contributes to both research and practice in requirements elicitation and specification.

\subsection{Research Implications}
ReqBrain contributes to requirements engineering research by demonstrating that fine-tuning LLMs can enhance the generation of high-quality requirements. It fills a gap in the largely manual process of requirements elicitation and specification by generating authentic and adequate requirements. Empirical validation using BERT and FRUGAL scores, along with human evaluations, underscores the effectiveness of customized LLMs over untuned models. 
Future research may extend our approach to cover further requirements-related tasks to advance the AI-assisted requirements generation approach, supported by our open-source dataset and methodology for continued collaboration and advancement in the field.

\subsection{Practice Implications}
Integrating ReqBrain into the requirements elicitation phase has the potential to improve the efficiency and accuracy of collecting, categorizing, and documenting requirements. Automatically generating authentic and adequate requirements may reduce manual workload and enable software engineers to focus more on strategic decision-making and stakeholder engagement. ReqBrain can be deployed individually or in group sessions. As an open-source model, it can be hosted locally to ensure data privacy and extended with RAG to process large volumes of text from sources like JIRA, databases, project documents, and interviews.

Although our empirical validation demonstrates that \mbox{ReqBrain} is effective in generating authentic and adequate requirements, human expert review remains essential to ensure ethical considerations, emotional intelligence, and contextual understanding.

\section{Threats to Validity}\label{sec:limitations}
In the following section, we will outline the various potential threats that could undermine the validity of our study.  

\subsection{Construct Validity} 
For the authentic construct, we used a two-step assessment. First, we deployed automated NLP-based metrics, then conducted a human evaluation. This order mitigates the limitations of automated NLP metrics in representing human alignment on clarity, coherence, relevance, realism, and implementability--key aspects of authenticity. Additionally, it helps reduce human evaluation across multiple low-quality models.  Further, as an intermediary step,  although this step is not replicable, between NLP-based metrics and human evaluators, the first author performed a manual review to confirm that these metrics identified the best-performing model. 
For the adequate construct, we employed human evaluations across four dimensions.

\subsection{Internal Validity}
Differences in participants’ interpretations of evaluation criteria may affect validity. To mitigate this, we selected experienced participants, held an onboarding session with detailed guidelines and knowledge refresher materials, and allowed participants to evaluate a few instances from each task after onboarding and ask questions. Anonymizing requirements in tasks B and C further minimized bias. 

Further, using the same ReqBrain-generated requirements in tasks B and C ensured that differences in evaluations were due to output quality and fine-tuning effectiveness rather than data variations. Requirements were anonymized and shuffled, and participants were asked to complete tasks in different orders to minimize carryover effects.

\subsection{External Validity}
Our sample sizes were two to three times larger than those determined by our a-priori power analyses, ensuring robust statistical power. 

We deem our four evaluators to have adequate job experience and education in the domain to provide a fair generalizability of their ratings. 
While this enhances the reliability of the insights, we acknowledge that four evaluators constitute a limited representation of cross-domain professionals dealing with requirements, which may limit generalizability. 

The evaluation for consistent with, missing from, and enhancing the overall completeness of, a given requirements specification dimensions in RQ2 was based on three distinct software projects coming from the same domain, which may result in not sufficiently representing the full diversity of software development projects. The present study may not wholly represent the real-world utility of ReqBrain: a case study aligned with the ISO 9241-11 definition of usability is underway to further explore this aspect.

\subsection{Conclusion Validity}
Given the ordinal nature of Likert scale responses, we used the median over the mean. We employed non-parametric statistical tests that do not assume normality or equal intervals. These choices ensure our statistical analyses are valid and appropriately aligned with the characteristics of our data.

\section{Conclusion}
We present ReqBrain, a fine-tuned large language model (LLM) and tool to generate authentic and adequate requirements supporting the AI-assisted requirements generation approach. 

To develop ReqBrain, we created an instruct dataset by manually reviewing and incorporating high-quality human-written requirements that are ISO 29148-compliant, ensuring that the training dataset reflects the real-world language and the nuanced complexities of industry requirements. Further, we fine-tuned five 7-billion-parameter models on the created instruct dataset.

ReqBrain, our fine-tuned variant of Zephyr-7b-beta, demonstrated superior performance with an 89.30\% F1 score using the BERT Score and 91.20 (out of 100) using the FRUGAL score, significantly outperforming other baseline fine-tuned models and the general-purpose ChatGPT-4o.

The human evaluation further confirmed that ReqBrain-generated requirements are authentic and indistinguishable from those written by humans, underscoring the model’s ability to produce outputs that meet industry standards in terms of clarity, coherence, relevance, realism, and implementability. Furthermore, ReqBrain demonstrates significant potential in generating adequate requirements that are ISO 29148-compliant, consistent with, missing from, and enhancing the overall completeness of a given requirements specification. 

ReqBrain can shorten the labor-intensive, manual process that is often prone to inconsistencies in traditional requirements development. It automates both the elicitation and specification phases by merging them into a single, efficient workflow, enabling the real-time, semi-automated creation of clear and actionable requirements instead of merely collecting raw data and then transforming it. This integration can reduce manual overhead and shorten iteration cycles while ensuring that the early requirements generated are authentic and adequate, thereby supporting requirements engineers in making strategic decisions such as accepting, rejecting, or modifying the generated requirements.

In summary, ReqBrain is a novel contribution, integrating fine-tuned LLMs into the requirements engineering process, marking a significant step toward an AI-assisted requirements generation approach. 

Future work will expand ReqBrain’s application in requirements engineering. We are currently conducting an industrial case study on ReqBrain’s usability in terms of efficiency, effectiveness, satisfaction, and process integration. Simultaneously, we are developing a RAG-enabled version to process company data, including logs, legacy documentation, and project records, to generate contextually accurate, ISO-compliant requirements. Additional work may broaden the instructional dataset to include acceptance criteria, use cases, user stories, and scenarios.

\bibliographystyle{IEEEtran}
\bibliography{references}

\end{document}